\documentclass[]{aa}  

\usepackage{graphicx}
\usepackage{txfonts}
\begin{document}

   \title{Direct Detection of the Tertiary Component in the Massive Multiple HD~150\,136 with VLTI.}

   \author{J. Sanchez-Bermudez\inst{\ref{inst1},\ref{inst2}}  \and R. Sch\"odel \inst{\ref{inst1}}  \and A. Alberdi\inst{\ref{inst1}} \and R. H. Barb\'a\inst{\ref{inst3}} \and C. A. Hummel\inst{\ref{inst2}} \and J. Ma\'iz Apell\'aniz\inst{\ref{inst1}}   \and J.-U. Pott\inst{\ref{inst4}}}

      \institute{Instituto de Astrof\'isica de Andaluc\'ia (CSIC), Glorieta de la Astronom\'ia S/N, 18008 Granada, Spain. 
              \email{joel@iaa.es}\label{inst1}
         \and
        European Southern Observatory, Karl-Schwarzschild-Stra$\beta$e 2, 85748 Garching, Germany. \email{sanchezj@eso.org}\label{inst2}
         \and
         Departamento de F\'isica, Universidad de la Serena, Benavente 980, 204000 La Serena, Chile.\label{inst3}
         \and
        Max-Planck-Institut f\"ur Astronomie, K\"onigstuhl 17, D-69117 Heidelberg, Germany. \label{inst4}}    

   \date{}
 
  \abstract
  {Massive stars are of fundamental importance for almost all aspects of astrophysics, but there still exist large gaps in our understanding of their properties and formation because they are rare and therefore distant. It has been found that most O-stars are multiples. It may well be that almost all massive ones are actually born as triples or higher multiples, but their large distances require milliarcsecond angular resolution for direct detection of the companions. }
  {HD~150\,136 is the nearest system to Earth with $>100\,M_{\odot}$, and provides a unique opportunity to study  an extremely massive system.  Recently, evidence for the existence of a third component in HD~150\,136, in addition to the tight spectroscopic binary that forms the main component, was found in spectroscopic observations. Our aim was to image and obtain astrometric and photometric measurements of this component using long baseline optical interferometry to further constrain the nature of this component.}
   {We observed HD150136 with the near-infrared instrument AMBER
attached to the ESO VLT Interferometer providing an angular resolution of 2 mas. The recovered closure phases are robust to
systematic errors and provide unique information on the source asymmetry.
Therefore, they are of crucial relevance for both image reconstruction
and model fitting of the source structure.}
   {The third component in HD~150\,136 is clearly detected in the high-quality data from AMBER. It is located at a projected angular distance of 7.3\,mas, or about 13\,AU at the line-of-sight distance of HD~150\,136, at a position angle of  209 degrees East of North, and has a flux ratio of $0.25$ with respect to the inner binary. Our findings are in agreement with Sana et al. (2013) and have permitted to improve the orbital solutions of the tertiary around the \textit{inner system}.}
   {We resolved the third component of HD~150\,136 in $J$, $H$ and $K$ filters.  The luminosity and color of the tertiary agrees with the predictions and shows that it is also an O main-sequence star. The small measured angular separation indicates that the tertiary may be approaching the periastron of its orbit. These results, only achievable with long baseline near infrared
interferometry, constitute the first step towards the understanding of the massive star formation mechanisms.}


   \maketitle
%

\section{Introduction}

The chemical composition of the universe cannot be fully understood without understanding the formation and evolution of high-mass stars. In every stage of the massive stars life cycle, they dominate the stellar feedback to the interstellar medium. They may both trigger and disrupt star formation, mainly, by the chemical enrichment produced by their strong stellar winds and the final supernova explosions that put an end to their lives. However, in spite of their importance, our knowledge about these objects and their evolution is still fragmentary. It is, principally, because they spend a significant part of their main-sequence lifetime while still embedded in their natal clouds, where the high extinction makes it difficult to observe their initial phases. 


One of the most striking features of massive stars is the high fraction of multiple systems among them. Most or all massive stars are suspected to be born as part of multiple systems \citep{Mason:2009fk,Maiz-Apellaniz:2010fk,Sana:2011kx}. Accurate knowledge of the multiplicity of massive stars is therefore key to understanding their formation and evolution. Our knowledge is especially fragmentary at the highest masses, where these objects are extremely rare.  Almost all known stars with masses $\geq50\,M_{\odot}$ are located at distances larger than one kiloparsec (kpc) and affected by large extinction, which makes them faint and hard to resolve. Because of these difficulties, the study of each individual extremely massive system matters in order to obtain reliable statistics of their properties.

\begin{figure}[ht]
\begin{center}
\includegraphics[width=6.5cm,clip]{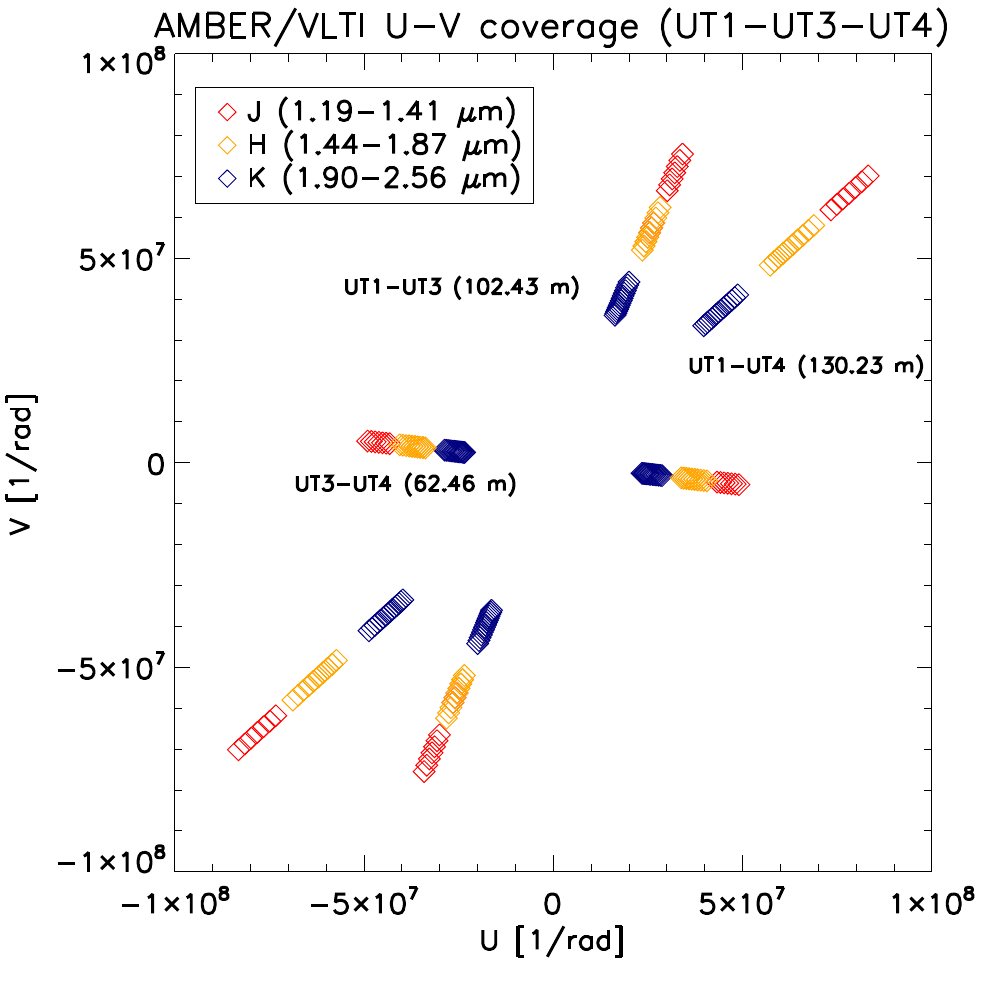}
\caption{$u-v$ coverage of our AMBER/VLTI observations of HD~150\,136. The baselines length is displayed in the plot}
\label{Fig:v2cp}
\end{center}
\end{figure}



HD~150\,136 in the open cluster NGC 6193 is such a system. \citet{Niemela:2005kx} found that the brightest component of HD~150\,136 is of spectral type O3. At a distance of $1.30\pm0.12$~kpc \citep{Herbst:1977uq} it is the closest star of this kind. In a detailed spectroscopic study \citet{Mahy:2012fk} found a third massive component in HD~150\,136, that was already tentatively predicted by \citet{Niemela:2005kx}. \citet{Mahy:2012fk} reported that the tightly bound inner system is composed of a primary ($P$) of spectral type O3V((f*))-3.5V((f+)) and a secondary ($S$) of O5.5-6V with eccentricity $e=0$, inclination $i=49^{\circ}\pm5^{\circ}$ and an orbital period of $2.67$ days The tertiary ($T$) was reported to be of spectral type O6.5-7V, and in orbit around the \textit{inner system} with a period of 3000 to 5500 days. During the refereeing process of this letter we became aware that \citet{Sana2013} determined the first solution of the orbit between $T$ and the \textit{inner system} using independent interferometric and spectroscopic data. They resolved $T$ with a projected angular separation of $\sim$9 milliarcseconds (mas) at a position angle of 236$^{\circ}$ N$\rightarrow$E with a period of 8.2 years and a eccentricity ($e$) of 0.73. The dynamical mass of HD~150\,136, found by those authors, agree with the previous predictions of \citet{Mahy:2012fk} with 63$\pm$10, 40$\pm$6 and 33$\pm$12 M$_{\odot}$ for $P$, $S$ and $T$, respectively.

The flux ratio between the \textit{inner system} and $T$ observed from the $H-$band interferometric measurements supports the previous estimation of spectral types and evolutionary stages (main-sequence stars) suggested by \citet{Mahy:2012fk}. In order to unveil the link of the observed multiplicity and the evolutionary models of massive stars, a continuing interferometric and spectroscopic monitoring  is mandatory.

Here, we report new long-baseline interferometric measurements of HD~150\,136 with the instrument AMBER at the ESO's Very Large Telescope Interferometer.\footnote{Based on observations collected at the European Organization for Astronomical Research in the Southern Hemisphere, Chile, within observing programme 090.D-0689(A).} Our aim was to resolve $T$ at different NIR frequencies in order to (i) provide accurate astrometric measurements to constrain the orbit of the system ($P$+$S$)+$T$ and (ii) to provide new photometric estimations of $T$ via direct measurement of its luminosity relative to the {\it inner system}.

\section{Observations and data reduction}

A single snapshot observation of HD~150\,136 was obtained with AMBER in its Low Resolution Mode (LR-HK), using the VLT unit telescopes (UTs) 1, 3, and 4 on March 4th, 2013 (JD 2456355.9). The triplet used for our observations has a maximum baseline length of 130\,m and a minimum baseline length of 63\,m. The synthesized beam obtained with this configuration has $3.59\times1.43$\,milliarcseconds (mas) with a position angle of $314.9^{\circ}$.  The instrumental setup allowed us to obtain measurements in the $J-$,$H-$, and $K-$bands with a spectral resolution of $R=\lambda/\Delta \lambda\approx35$.  A standard calibrator -- science target -- calibrator observing sequence was used. The chosen calibrator, HD~149\,835, is separated by $1.47^{\circ}$ deg from HD~150\,136. It is a K0III star with magnitude of $J=5.5$, $H=5.1$, and $K=4.9$, similar to the corresponding magnitudes of the target (see Table \ref{apmag}). The airmass reported for the calibrator is $1.19$, and $1.15$ for the target, respectively.  Fig.\,\ref{Fig:v2cp} shows the $u-v$ coverage of our observations.

\begin{figure*}[ht]
\begin{minipage}[ht]{11.5cm}
\begin{center}
\hspace{-5mm}
\includegraphics[width=10 cm,clip]{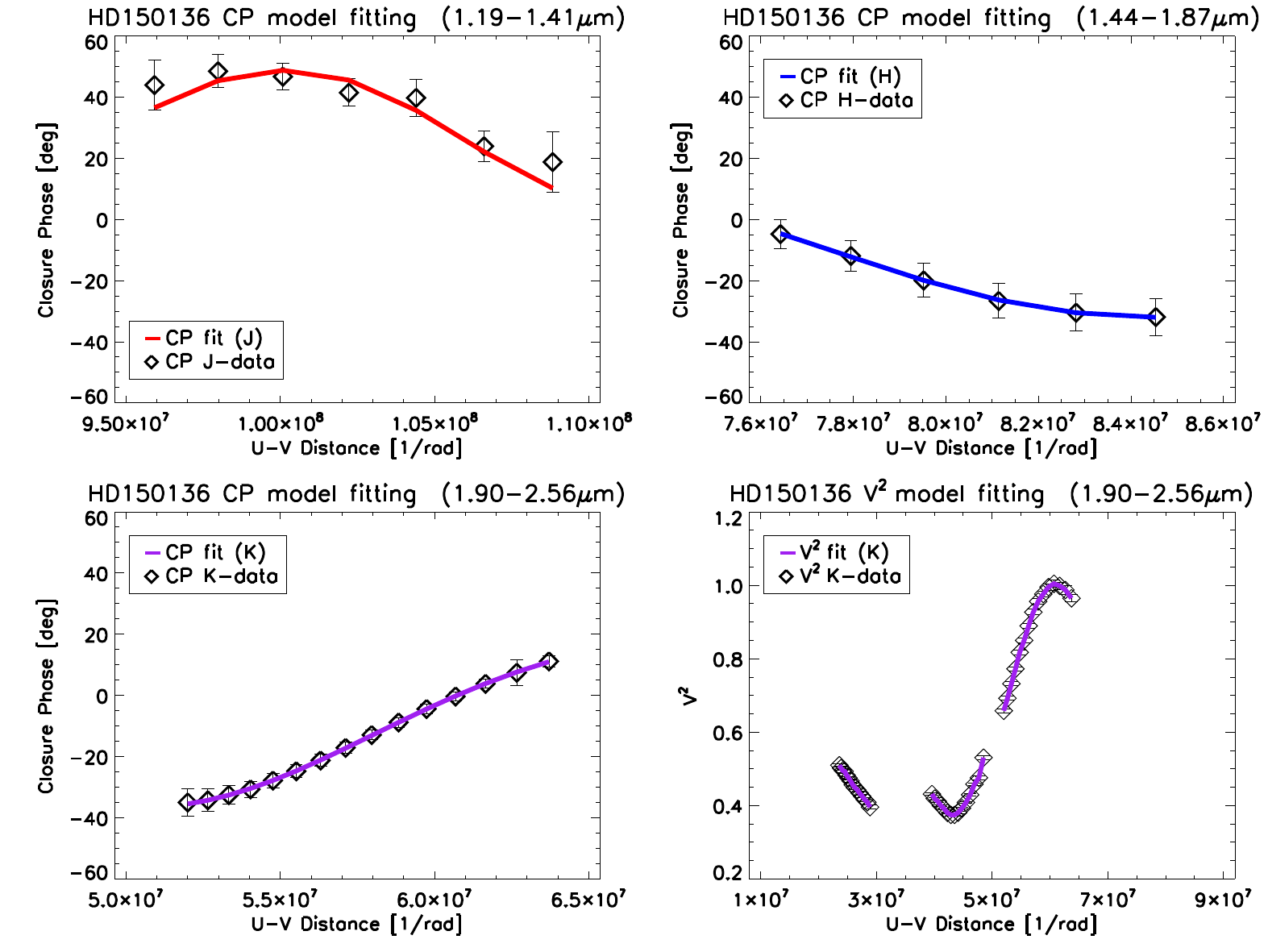}
\end{center}
\end{minipage}
\begin{minipage}[ht]{7.2cm}
\begin{center}
\includegraphics[width=7.2cm,clip]{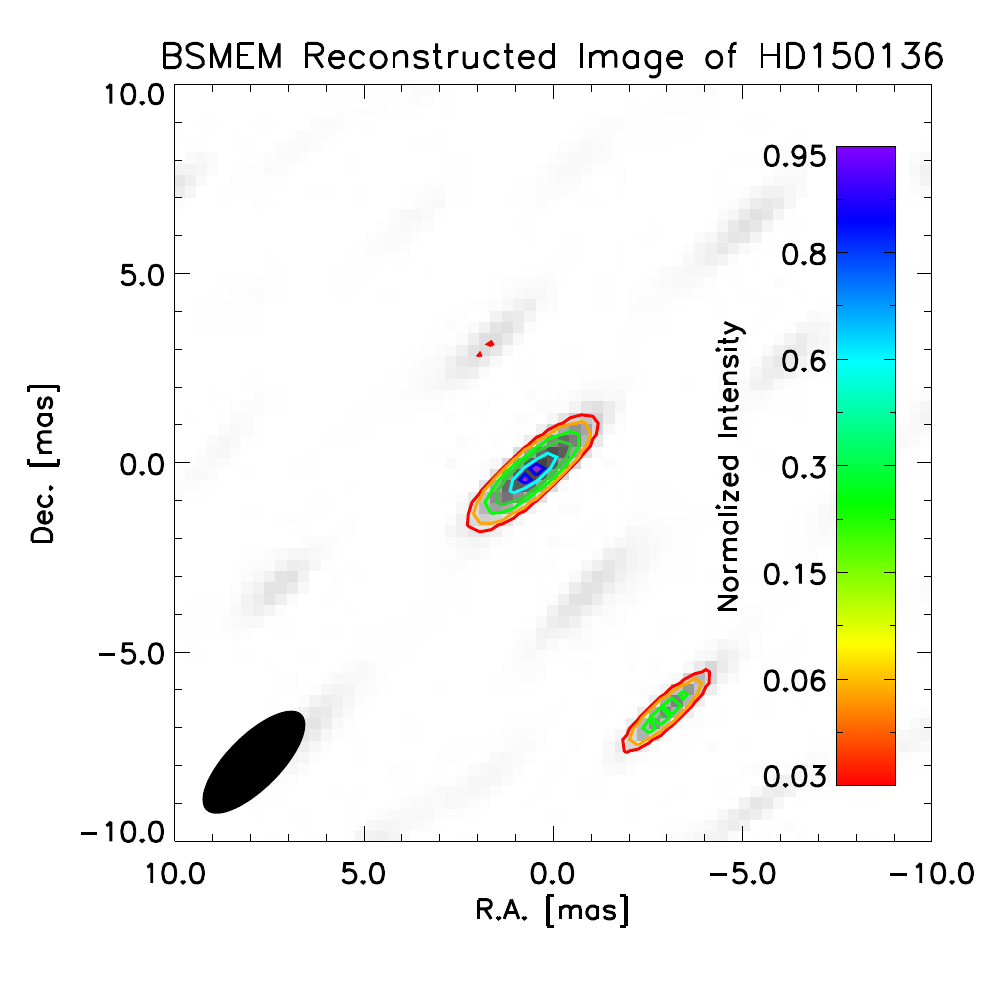}
\end{center}
\end{minipage}
\caption{{\bf a) Left:} CP and $V^2$ Model fitting. Data are shown in black diamonds and the best obtained model in colored lines. {\bf b) Right:} BSMEM reconstructed image of HD~150\,136. The total flux is normalized, contours represent 3, 7, 15, 30, 60, 80 and 95\% of the maximum. The black ellipse at the bottom-left represents the synthesized beam of our interferometer.}
\label{Fig:cpfit}
\end{figure*}

For the AMBER data reduction we used the \textit{amdlib3}\footnote{Available at: http://www.jmmc.fr/amberdrs}  data reduction software, which uses the algorithms by \citet{Tatulli:2007ve} and \citet{Chelli:2009qf}. In order to eliminate frames which were deteriorated by variable atmospheric conditions and/or technical problems (e.g., shifts in the path delay), we selected only the 20\% of frames with the highest signal-to-noise ratio (SNR). The two sets of calibrator observations exhibit similar $V^2$ response within 5\% accuracy. Hence, we interpolated a linear fit to the calibrator visibilities for the epoch of the science target to normalize our visibilities. The obtained calibrated squared visibilities ($V^2$) at $K-$band and closure phases (CPs) are displayed in Fig.\,\ref{Fig:cpfit}. 

\section{Analysis}

We note that the apparent separation between $P$ and $S$ in the {\it inner system} of HD~150\,136 is about $0.1$\,mas \citep[][]{Mahy:2012fk}, and thus more than one order of magnitude smaller than what can be resolved with VLTI in the near-infrared (NIR). Therefore, to fit the calibrated measurements we chose the model of a binary, composed of two unresolved sources ({\it inner system} plus tertiary companion, $T$). The model fitting was done with the LITpro software  to obtain the best-fit model parameters \citep{Tallon-Bosc:2008zr}.




When fitting the CPs and $V^2$ of all bands simultaneously, we noted significant systematic residuals at the shorter wavelengths: the $V^{2}$ values were consistently  too low for all baselines in $J$ and showed some systematic errors for the shortest 
baseline (UT1-UT3) in $H$. Since the photon count of the target was unexpectedly low in $J$ and $H$, we suspect some technical problems were present in the observations, but could not identify the exact cause. Hence, to circumvent the systematics observed in the $V^2$ from the unsatisfactory calibration in $J-$ and $H-$band, we chose to restrict the model fit only to the CP data of the three bands. Since CP data by definition eliminate with high reliability most of the atmospheric effects, they are, usually, much more robust to systematic errors than $V^2$, which are strongly affected by weather conditions (e.g. strong variable seeing).  Therefore, the use of CP to our model fitting provides a good estimator of the real brightness distribution of the source. The model was fitted to the combined CPs from all bands and to the CPs  from each band individually. The resulting fits for the individual bands are shown in Fig.\,\ref{Fig:cpfit}. Any biases are small with respect
to the dynamic range of the data.

The best-fit parameters are listed in Table\,\ref{params}. We find an angular separation of $7.27\pm0.05$\,mas of $T$
from the {\it inner system}, a Position angle on the sky of $209\pm2^{\circ}$ East of North, and a flux ratio of $f_{T}/f_{inner}=0.25\pm0.03$. The $1\,\sigma$ uncertainties were estimated from the standard deviations of the best-fit values from the fits to the individual bands. We should note that in the case we extend the modelfit to both the $V^{2}$ amplitudes
and the CPs for the three individual bands, the resulting fits agree within 
the $1\,\sigma$ uncertainties with the CP-only fits. This gives extra support to our
results. For illustrative purposes,  we show in Fig.\,\ref{Fig:cpfit} the fit to the $K-$band 
$V^{2}$ data of the reported model. 


Image reconstruction from the interferometric data was performed with the BSMEM package \citep{Lawson:2004uq}. This code uses a maximum entropy algorithm to recover the real brightness distribution of the source. For the image reconstruction we decide to include all CP and $V^2$ of the three observed wavelengths in order to improve the quality of the image and reduce the sidelobes. The best reconstructed image was created after 45 iterations. It is displayed in Fig.\,\ref{Fig:cpfit} and it is consistent with the parameters obtained from the model fitting.

\begin{table}[ht]
\caption{Best-fit parameters of a binary model (inner system plus tertiary) for HD~150\,136  to the closure phases to all bands combined and to each band individually. \label{params}} 
\centering  
\begin{tabular}{l r r r r } 
\hline \hline                        
Parameter & Combined &$J$ & $H$ & $K$ \\ 
\hline                  
$f_{inner}$\tablefootmark{a}         &  $0.80$  & $0.78$  &   $0.80$  &  $0.82$ \\
$f_{T}$\tablefootmark{b}              & $0.20$  & $0.22$  &   $0.20$  &  $0.18$  \\
$d$ [mas]\tablefootmark{c}         &  $7.27$  & $7.27$  &   $7.19$  &  $7.19$ \\
$\Phi$ [deg]\tablefootmark{d}     &  $209.0$  & $210.2$  &   $206.7$  &  $210.7$ \\
\hline
\end{tabular}
\tablefoot{
\tablefoottext{a}{Fraction of total flux contained in the inner system ($P$+$S$).}
\tablefoottext{b}{Fraction of total flux contained in the tertiary ($T$).}
\tablefoottext{c}{Angular separation between inner system and $T$ in milliarcseconds.}
\tablefoottext{d}{Projected angle of the system on the sky.}
}
\end{table}

\section{Results and discussion}


The measured flux ratio between the tertiary and the inner system  agrees well with the flux ratio that results from the $V$ magnitudes computed for the three components by \citet{Mahy:2012fk}: $f_{theor, T/inner} = 0.19\pm0.08$, and with the $H-$band astrometric measurements by \citet{Sana2013}: $f_{PIONIER, T/inner} = 0.24\pm0.02$. For a more detailed analysis of the source brightness, we used the Bayesian code CHORIZOS \citep{Maiz2004} to obtain the extinction for HD~150\,136 using as input the Strömgren + NIR photometry in Table\ref{apmag}. More specifically, we used the Milky Way SED grid of \citep{Maiz2012a} and the new family of extinction laws described in  \citep{Maiz2012b}. The results of the fit are a monochromatic color excess $E(4405-5495) = 0.431\pm0.009$ and an extinction law with $R_{5495} = 4.12\pm0.13$. The somewhat large value of $R_{5495}$ is typical of stars in H\,{\sc ii} regions such as HD~150\,136 and the value of $E(4405-5495)$ is within the expected range for a star of its Galactic coordinates and distance. The above values correspond to $A_J = 0.560\pm 0.021$, $A_H = 0.360\pm 0.013$ and $A_K = 0.230\pm 0.009$. Note, however, that the largest residual of the fit comes from the $K$-band photometry, suggesting a small excess in the observed spectrum of the order of 0.03 magnitudes. We also verified that the existing Tycho-2 photometry agrees with the model SED derived from CHORIZOS.

\begin{table}[ht]
\caption{Str\"omgren and NIR Photometry of HD~150\,136
. \label{apmag}} 
\centering  
\begin{tabular}{l r||lr } 
\hline \hline                        
\multicolumn{2}{c||}{Str\"omgren Photometry\tablefootmark{a} } & \multicolumn{2}{c}{NIR Photometry\tablefootmark{b}} \\
\hline                  
$V$ &5.647$\pm$0.001 &$ J$ & 5.15$\pm$0.037 \\
$b-y$ &0.192$\pm$0.009 & $H$ & 5.09$\pm$0.018 \\
$m_1$ &-0.043$\pm$0.011 & $K$ & 4.99$\pm$0.018 \\
$c_1$  &-0.114$\pm$0.002 &  &  \\
\hline
\end{tabular}
\tablefoot{
\tablefoottext{a}{Values taken from \citet{Gronbech1976}}
\tablefoottext{b}{$J-$band values obtained from \citet{Clarke2005} and $H-$ and $K-$band values from the 2MASS catalog \citep{Skrutskie2006} }
}
\end{table}

\begin{figure}[ht]
\begin{center}
\includegraphics[width=6cm]{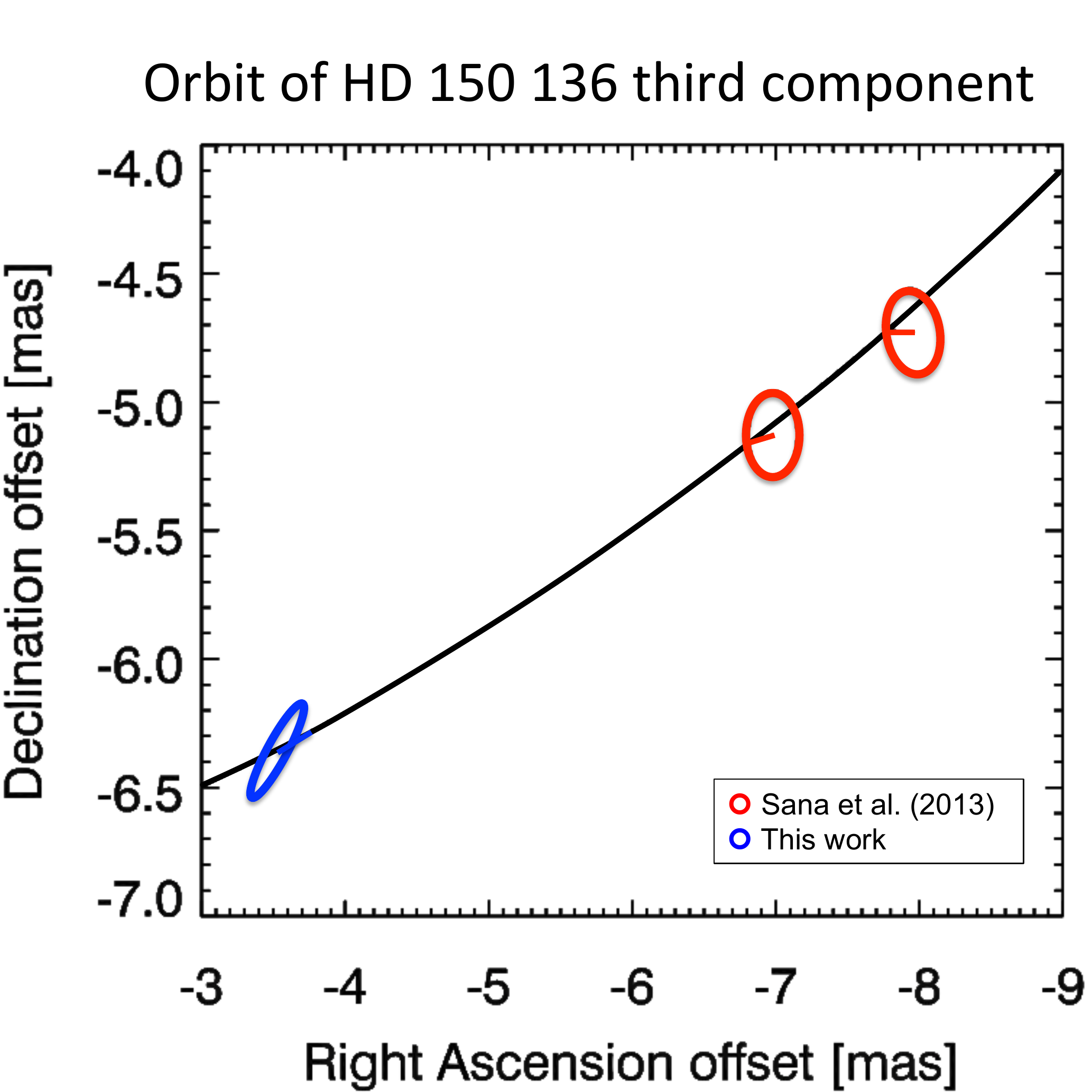}
\caption{Orbital motion of $T$ around the \textit{inner system}. The coordinate system is centered on the $P+S$ position, the red ellipses correspond to the astrometric positions of \citet{Sana2013} and the blue one to our AMBER data. The size of the ellipse is equivalent to 1 $\sigma$ uncertainty and the straight blue and red lines indicate the
offsets of the measured astrometric positions from the best orbital fit. The continuos black line represents the best fit to the orbital motion.}
\label{Fig:orbitfig}
\end{center}
\end{figure}

The age of the HD~150\,136 system cannot be too young ($\sim 0.1$ Ma) because the system has clearly emerged from the embedded phase. On the other hand, it cannot be too old ($\sim 2$ Ma or older) because in that case the more massive component would have evolved to a later spectral type than O3 V. Therefore, the age must be close to 1 Ma. Using the corresponding Geneva isochrone without rotation \citep{Lejeune2001} and assuming $T_{\rm eff}$ of 44\,500 K, 39\,000 K, and 37\,000 K for $P$, $S$, and $T$, respectively, we derive a flux ratio $T$/($P$+$S$) in the $K$ band of 0.22, which is within one sigma of our value. Such a triple system should have a combined absolute magnitude in the $K$ band of $-5.04$ (as derived from the assumed isochrone and temperatures). On the other hand, after correcting for a distance of 1.3 kpc and the extinction derived above , a measured $K$ magnitude of 4.991 results in an absolute magnitude $M_K = -5.81$, significantly brighter than expected. If, however, we assume an age of 1.8 Ma the total expected magnitude of the system would increase by more than a magnitude. Hence, one possibility is that the system is several hundred thousand years older than 1 Ma.

From the flux ratio between the \textit{inner system} and $T$ obtained at each filter as well as the described photometric analysis, we obtained the following absolute magnitudes for $T$: $M_{T, J}=-4.34\pm0.23$, $M_{T, H}=-4.09\pm0.21$ and $M_{T, K}=-3.93\pm0.21$. These values are in agreement (within $1\sigma$ confidence) with the colors of an O6.5-O7 main-sequence star according to the calibration of O-type stars developed by \citet{Martins2006}. Consistently, the $J-$, $H-$ and $K-$ absolute magnitudes of the \textit{inner system} ($M_{inner, J}=-5.71\pm0.23$, $M_{inner, H}=-5.60\pm0.21$ and $M_{inner, K}=-5.59\pm0.21$) are also (within $1\sigma$ accuracy) in agreement with a combined pair of O3+O5.5 stars ($M_{theor, J}=-5.63$, $M_{theor, H}=-5.52$ and $M_{theor, K}=-5.42$; \citet{Martins2006}), as it was expected. Such a photometric analysis, only feasible with our multi-wavelength interferometric observations, clearly provide us consistent information on the nature of HD~150\,136. 

Additionally, we find an angular separation of $T$ from the \textit{inner system} of $7.22\pm0.22$\,mas and a position angle of $209\pm2^{\circ}$ N$\rightarrow$E. Those measurements are consistent with the results obtained by \citet{Sana2013}. Nevertheless, in order to improve the knowledge of $T$ orbital motion, we performed a model fitting of the period and epoch (keeping the rest of the orbital parameters fixed) of $T$ towards the \textit{inner system}. For this purpose we combined our new interferometric measurement with the radial velocities and the two astrometric epochs of \citet{Sana2013}. Our best fit retrieves a change in the orbital period from 3008 to 2770 days and a modification in the epoch (periastron passage) from 2451241 to 2451614. A preliminary analysis of our  spectroscopic data from monitoring HD~150\,136 within the \textit{OWN Survey project} \citep{Barba:2010fk} from April 2005 to July 2012, and the recent published spectral analysis of \citet{Sana2013} indicates that $T$ may indeed be approaching the periastron of its orbit within the next two years. Interferometric measurements at such position will help us to derive, without ambiguity, the semi-minor axis of the orbit and to constrain with high precision the other orbital parameters. Figure \ref{Fig:orbitfig} displays the best fit of our interferometric data in addition to the previous measurements.


HD~150\,136 is one more example of the increasing number of O stars that belongs to multiple systems with gravitational bound components at different spatial scales (e.g. Herschel\,36; \citet{Arias2010}). The fact that massive stars are born in multiple systems have strong implications on their star formation scenarios. Thus, the orbital motion of component $T$ around the HD~150\,136 \textit{inner system} deserves a complete study in order to test the coplanarity of the orbits in this massive multiple, which can provide us with important clues about its formation. The currently favored theoretical models for the formation of high-mass stars are: (i) the collapse of a massive monolithic protostellar core \citep{Krumholz2009} and (ii) competitive accretion in clusters \citep{Bonnell2006}. On average, the components of a massive multiple that forms through collapse of a monolithic cloud are expected to have coplanar orbits, while the formation of massive stars from clouds with hierarchical sub-structure (or posterior formation of multiples from dynamical encounters) will favor the creation of systems with more randomly distributed orbits. High resolution spectroscopy will have to be combined with astrometry from the high angular resolution ($\sim$2 mas) of AMBER NIR interferometry to acquire the necessary data for such future work.


\begin{acknowledgements}
We thank the referee for his/her useful comments. JSB, RS and AA acknowledge support by grants AYA2009-13036, AYA2010-17631
and AYA2012-38491-CO2-02 of the Spanish Ministry of Economy and Competitiveness,
  and by grant P08-TIC-4075 of the Junta de Andaluc\'ia. RS
  acknowledges support by the Ram\'on y Cajal programme of the Spanish
  Ministry of Economy and Competitiveness. JMA acknowledges support by grants AYA2010-17631 and AYA2010-15081 of the Spanish Ministry of Economy and Competitiveness. RHB acknowledges financial support from FONDECYT Regular Project No. 1120668. JSB acknowledges support by the ESO studentship program, to the JAE-PreDoc program of the Spanish Consejo Superior de Investigaciones Cient\'ificas (CSIC) and to CAH for the OYSTER\footnote{Available at: http://www.eso.org/$\sim$chummel/oyster/oyster.html} software.
\end{acknowledgements}

\bibliographystyle{aa}
\bibliography{/Users/joel/Documents/AMBER_2012/HD150136/Joel_HD150136/HD150136_2}

\end{document}